**Electrical control and transport of tightly bound interlayer excitons in a $MoSe_2$/hBN/$MoSe_2$ heterostructure**

Lifu Zhang[1,*], Liuxin Gu[1,*], Ruihao Ni[1], Ming Xie[2], Suji Park[3], Houk Jang[3], Rundong Ma[1], Takashi Taniguchi[4], Kenji Watanabe[5], You Zhou[1,6,†]

[1]Department of Materials Science and Engineering, University of Maryland, College Park, MD 20742, USA

[2]Condensed Matter Theory Center, University of Maryland, College Park, MD 20742, USA

[3]Center for Functional Nanomaterials, Brookhaven National Laboratory, Upton, NY 11973, USA

[4]Research Center for Electronic and Optical Materials, National Institute for Materials Science, 1-1 Namiki, Tsukuba 305-0044, Japan

[5]Research Center for Materials Nanoarchitectonics, National Institute for Materials Science, 1-1 Namiki, Tsukuba 305-0044, Japan

[6]Maryland Quantum Materials Center, College Park, Maryland 20742, USA

†To whom correspondence should be addressed: youzhou@umd.edu

*These authors contributed equally to this work.

## Abstract

Controlling interlayer excitons in van der Waals heterostructures holds promise for exploring Bose-Einstein condensates and developing novel optoelectronic applications, such as excitonic integrated circuits. Despite intensive studies, several key fundamental properties of interlayer excitons, such as their binding energies and interactions with charges, remain not well understood. Here we report the formation of momentum-direct interlayer excitons in a high-quality $MoSe_2$/hBN/$MoSe_2$ heterostructure under an electric field, characterized by bright photoluminescence (PL) emission with high quantum yield and a narrow linewidth of less than 4 meV. These interlayer excitons show electrically tunable emission energy spanning ~180 meV through the Stark effect, and exhibit a sizable binding energy of ~81 meV in the intrinsic regime, along with trion binding energies of a few millielectronvolts. Remarkably, we demonstrate the long-range transport of interlayer excitons with a characteristic diffusion length exceeding 10 μm, which can be attributed, in part, to their dipolar repulsive interactions. Spatially and polarization-resolved spectroscopic studies reveal rich exciton physics in the system, such as valley polarization, local trapping, and the possible existence of dark interlayer excitons. The formation and transport of tightly bound interlayer excitons with narrow linewidth, coupled with the ability to electrically manipulate their properties, open exciting new avenues for exploring quantum many-body physics, including excitonic condensate and superfluidity, and for developing novel optoelectronic devices, such as exciton and photon routers.

Spatially indirect excitons (IXs) with long lifetimes and electric dipole moments can form macroscopic coherent quantum phases, such as exciton superfluidity, at high densities and low temperatures[1-6]. The quantum degeneracy temperature of IXs, which scales proportionally to the exciton binding energy, is limited in traditional III-V double–quantum well systems[6-8]. Transition metal dichalcogenide (TMD) heterostructures, predicted to host tightly bound IXs[6], have recently emerged as an exciting new platform for exploring coherent high-temperature condensate and superfluid states[9-13]. Furthermore, these IXs exhibit spin-valley coupling[14-16], strong dipolar interactions[17, 18], and electrical tunability via the Stark effect[19-21], thereby offering a system rich for exploring exciton physics.

To date, optical studies of interlayer excitons have primarily focused on multilayer TMD hetero- and homo-junctions[15-33]. In heterostructures, local variations in the twist angle and strain can induce substantial inhomogeneous broadening of IXs, for instance, by lattice reconstruction, introducing a disorder potential atop the periodic moiré lattice[23, 24, 26, 27, 34, 35]. On the other hand, although disorder potential can be greatly reduced in natural homo-layered TMDs, these systems become momentum-indirect[21, 28, 29, 36], which feature significant non-radiative broadening of IX. As a result, IXs in TMDs typically manifest much larger linewidths than their intralayer counterpart in monolayers, which hinders the formation of exciton superfluidity and limits our understanding of their critical properties, such as their binding energies and their interactions with free carriers.

Here, we employ a homo-bilayer TMD system, with two $MoSe_2$ monolayers separated by an atomically thin layer of hBN. The thin hBN spacer preserves the direct bandgap in bilayers while allowing for carrier tunneling and IX emission. Although IX PL has been observed in similar structures, earlier works focused on the hybridization of IXs and intralayer excitons[37, 38]. In this work, we demonstrate a low-disorder system with degenerate intralayer exciton energies and sharp IX PL linewidth. The high optical quality allows us to directly measure the interlayer exciton and trion binding energies and observe the long-range transport and trapping of IXs, creating a new platform for studying exciton transport, photophysics, and condensates.

We first establish the electric-field and doping control of interlayer excitons in a $MoSe_2$/hBN/$MoSe_2$ device. We align the two monolayers close to zero degree to promote the carrier tunneling and IX emission (**Fig. 1a,** see **Table S1** for hBN thickness in different devices). Under zero electric field, the sample's PL spectrum at 6 K is dominated by intralayer neutral exciton $X_0$ and charged excitons $X_T$ (**Fig. 1b**). The two monolayers have degenerate $X_0$ energies, suggesting minimal strain difference. Next, we apply an electric field while keeping both layers intrinsic ($V_{BG} = -\alpha V_{TG}$, where $\alpha$ is the thickness ratio of the top and bottom hBN layers, **Fig. S1**). Under a large electric field, an additional peak, IX, emerges at a lower energy of 1.596 eV with a narrow linewidth of ~3.7 meV, accompanied by a reduction in $X_0$ and $X_T$ intensities. Remarkably, the quantum yield of IX is comparable to that of the intralayer excitons in monolayer $MoSe_2$ under zero electric field (**Fig. S2c**). The electric-field dependent PL spectra in **Figs. 1c** show a clear Stark effect of IX, from which we estimate a permanent out-of-plane dipole of $u = e * 0.81\ nm$ for IX (SI [39]), consistent with the thickness of the hBN spacer (two layers) plus the finite thickness of $MoSe_2$. From reflectance measurements, we verify that both layers are intrinsic, and the $X_T$ emission likely originates from in-gap states in $MoSe_2$ (**Fig. S2a**).

We attribute IXs to the momentum-direct interlayer exciton at the K-K transition and extract their binding energies. In particular, we calculate the IX energy under zero electric field to be 1.731 eV from the Stark effect, which is given by the difference between the quasiparticle bandgap and the interlayer binding energy. By measuring the Rydberg states of $X_0$, we estimate the quasiparticle bandgap of monolayer $MoSe_2$ to be ~1.812 eV using a screened Keldysh potential model[40-42] (SI & **Fig. S2a** [[39]]). Therefore we estimate the binding energy of IX to be ~81 meV, consistent with theoretical calculations[6, 43] (see **Table S1** for its dependence on hBN thickness). We note that, different from other studies[30, 37, 44], we did not observe clear signatures of IX or their anti-crossings with $X_0$ in reflectance (**Fig. S2b**), suggesting their low oscillator strength and weaker carrier tunneling.

We further investigate the doping dependence of IX. Under symmetric gating with $V_{BG} = \alpha V_{TG}$, the measured doping-dependent PL resembles that of monolayers without any IX emission, as the carriers are evenly distributed across the two layers (**Fig. S3a**). To make IX energetically favorable and detectable through PL, we vary the total doping under a large electric field, following $V_{BG} = \alpha V_{TG} + \delta$ (**Fig. 1d**). The $X_0$ intensity reduces upon doping but remains finite. This is because the applied field polarizes carriers into one layer and makes the other layer intrinsic, which is further verified by measuring $X_0$ reflectance (**Figs. 1e, S2d, and S4**). In contrast, the intrinsic $IX_0$ vanishes upon doping one layer, with an additional peak appearing below it, which we identify as charged interlayer excitons, i.e., interlayer trions or polarons ($IX_T$ in **Fig. 1d**). The trion binding energy (in the low doping regimes) is estimated to be ~3 meV for negatively charged and ~5 meV for positively charged IXs, with the difference likely due to the varying effective masses of electrons and holes ($m_e^* > m_h^*$)[45, 46].

The charged IX shows a significant blueshift with increasing total gate voltages on both electron and hole sides (also in **Fig. S3b-d**). In such a bilayer with layer-polarized carriers, varying the gate voltage alters both the doping level and the electric field between layers. The effective screening by the highly doped layer and weak screening by the intrinsic layer leads to a reduction in the electric field magnitude and blueshift of IX with increasing voltage (see **Fig. 1f** and Discussion III [[39]]). To separate doping and electric-field influences on this blueshift, we measure the field-dependent PL of charged IXs at various doping levels and extract their energies at zero field, from which we find that the blueshift is primarily due to the electric-field effect (**Fig. S5**).

We then investigate the valley polarization properties of IX by exciting the sample with a circularly polarized laser at 1.95 eV and detecting the preservation of circular polarization in PL. Intriguingly, IX exhibits a ~ 20% degree of circular polarization (DOCP), defined as $(I_{co} - I_{cross})/(I_{co} + I_{cross})$, while the DOCP of intralayer excitons is almost zero (**Figs. 2a** and **S6**). In $MoSe_2$, the absence of intralayer excitons DOCP is due to rapid valley mixing caused by electron-hole exchange[47-51]. In contrast, the electron-hole exchange of IX can be much weaker thanks to the spatial separation of the electron-hole wavefunction[15, 52, 53]. The finite DOCP also suggests that the formation of IXs occurs faster than intervalley electron-hole exchange. The DOCP of IX does not vary significantly with the electric field except in the crossover region, where it hybridizes with intralayer species, as expected (**Fig. 2b**).

Next, we examine the optical nonlinearity of IX by measuring their PL spectra under different excitation powers (**Fig. 2c**). With higher power, IX blueshifts while no obvious shift of $X_0$ and $X_T$ can be observed (**Figs. S7**). The stronger nonlinearity of IX compared to intralayer excitons arises from their repulsive dipole-dipole interactions due to the aligned electric dipols[17, 21, 25]. The *n*-doped IX exhibits a similar trend but with a larger blueshift at the same excitation powers (**Fig. S7**). In the hole-doped case, however, we observe a shift in the $X_T$ energy, which suggests optical pumping may alter the doping levels (**Fig. S8**), likely due to Auger-assisted hole tunneling across hBN dielectrics[54].

We focus our analysis of the power-dependent PL on the intrinsic and $n$-doped regimes to avoid photo-doping effects. From fitting, we find a ~3.1 meV blueshift of neutral IX and ~4.1 meV blueshift of $n$-doped IX along with linewidth broadening of ~2-3 meV when the excitation reaches 16 μW (**Fig. 2d**). The IX density, $n_{IX}$, can be estimated using the mean-field parallel plate-capacitance model[20, 55] following: $\delta E = e u n_{IX}/\varepsilon$, where $\delta E$ is the energy shift, $e$ is the electron charge and $\varepsilon$ is the dielectric constant (Discussion II and IV [[39]]). Notably, we estimate a maximum exciton density. $n_{IX}$ of ~$7.6 \times 10^{10}$ cm$^{-2}$ and ~$1.3 \times 10^{11}$ cm$^{-2}$ under our highest excitation power of 16 μW and 64 μW, for the intrinsic and $n$-doped case, respectively. The integrated IX PL intensity has a linear dependence on $n_{IX}$ in the $n$-doped case and has a nearly quadratic dependence in the intrinsic region (**Fig. S7d**). This suggests that radiative recombination in $n$-doped IX is relatively independent of exciton density, whereas in the intrinsic region, the recombination can be density-dependent due to processes like collective radiation, biexciton formation, and Auger recombination[36, 56, 57].

The electrical control of spectrally sharp IXs with strong dipolar interactions in our device allows us to probe the transport properties of IX. **Figures 3a, 3b, and S9** show the spatial map of IX PL under different excitation powers. With the excitation laser fixed near the sample's bottom right corner, the IX PL is visible away from the laser spot and extends further with increasing excitation power, showing distinct local bright areas. In contrast, no long-range transport is observed for intralayer excitons, even at the highest power (**Fig. S9c**).

To quantify the IX transport, we extract the normalized, radially averaged PL intensity $I_{norm}(r)$ from these spatial maps, where $r$ is the distance from the center of the laser spot. Since spatial inhomogeneities significantly affect diffusion profiles, we focus on data collected from another sample area with a more uniform IX distribution (**Figs. 3c and S10**). Exciton density near the laser resembles the Gaussian beam profile and asymptotically approaches $n_{IX} \propto e^{-r/L_D}/\sqrt{r/L_D}$ further away (see Discussion IV [[39]]). Fitting the density profile away from the laser (dashed lines in **Fig. 3c**), we extract a diffusion length $L_D = 17 \pm 6$ μm for the neutral IX, which does not vary significantly with excitation power. As we move closer to the laser and increase the laser power, the interactions-driven current becomes increasingly prominent, surpassing the pure diffusion current driven by the density gradient (**Fig. S11** and Discussion IV [[39]]). It is important to mention that we have ignored other nonlinear exciton decays such as exciton-exciton annihilation[32, 58] and exciton-phonon interactions[59], since we focus on regions away from the laser excitation. However, these nonlinear processes, along with the drift and diffusion of IX, can contribute to the measured sublinear power dependence of exciton densities (**Fig. S7e**).

The bright spots away from the laser in the diffusion maps can act as a natural trap for IX, which increases the local exciton density and the condensate temperature. **Figure S12** show the PL of neutral and doped IXs collected from a local bright spot ~2.9 μm away from the laser, with a sharp linewidth of ~4-5 meV similar to **Fig. 1b**. A blueshift of ~1 meV, along with linewidth broadening of ~1 meV, is observed under the maximum pump power (**Figs. S12c and S12d**). We estimate the locally trapped exciton density to be ~$10^{10}$ cm$^{-2}$, only a few times less than that under direct laser excitation (**Fig. 3d**).

Lastly, we present the intriguing observation of the periodic modulations of $X_0$ and $X_T$ PL intensities in response to varying electric fields under fixed doping (**Fig. 4a**). Under positive fields, electrons become layer-polarized, resulting in predominantly emission from the top layer's *n*-doped trion, $X_{Tt}$, and the bottom layer's neutral exciton, $X_{0b}$. IX emerges when its energy drops below $X_{Tt}$, different from the intrinsic and *p*-doped cases, where IX appears when it becomes degenerate with $X_0$ (**Figs. 1** & **S13**). This difference stems from the more efficient hole tunneling than electrons in TMD heteorstructures[9, 38, 54, 60, 61]: hole tunneling leads to IX-$X_{Tt}$ hybridization in the *n*-doped region (**Fig. 4b**), as opposed to IX-$X_{0b}$ coupling in the *p*-doped and intrinsic cases (**Fig. 4c**).

Increasing electric fields typically induce a continuous change in carrier density in each layer and a monotonic increase (decrease) in $X_{Tt}$ ($X_{0b}$) intensity. However, at three distinct electric fields, we observe reduced $X_{Tt}$ and enhanced $X_{0b}$ emissions (**Figs. 4d and S13b**). These anomalies, marked as $I_e$, $II_e$, and $III_e$, recur periodically at ~21 mV/nm intervals. Feature $III_e$ appears at the electric field where IX becomes degenerate with neutral exciton $X_{0b}$. This energy degeneracy can lead to a resonantly enhanced IX-$X_{0b}$ coupling, introducing an additional decay pathway for the long-lived IX into $X_{0b}$, which enhances $X_{0b}$ and reduces $X_{Tt}$ emission (**Fig. S13**). Similarly, features $I_e$ and $II_e$ may arise when possible dark interlayer excitons with lower energy than IX reach degeneracy with $X_{0b}$. **Figure 4e** shows a map of integrated $X_0$ intensity as we vary top and bottom gates, where we observe similar anomalous features in the intrinsic and *p*-doped regimes. Across all doping conditions, the energy difference between bright and dark excitons is ~17 meV. The dark excitons may emerge due to misaligned $MoSe_2$ layers creating local atomic registry variations. This moiré superlattice could give rise to IX species with different optical selection rules and oscillator strengths[37, 62]. Alternatively, interlayer electron–phonon coupling might generate phonon replicas of IXs, a hypothesis supported by the similar energy scales to phonon energies in hBN and $MoSe_2$[63-66]. Indeed, in a device with a larger twist angle (~3°), no IX emission is observed but the modulation of $X_0$ and $X_T$ is still present, likely due to their coupling with momentum-indirect dark IXs. Detailed optical and transport investigations will be crucial to clarify the dynamics between dark and bright IXs.

The electrically tunable IX with sharp linewidth forms a promising platform for exploring many-body excitonic physics such as condensate and superfluidity. Using the experimentally extracted $n_{IX}$ in the intrinsic regime, we estimate the quantum degeneracy temperature , $T_q = \frac{2\pi\hbar^2 n_{IX}}{m^* k_B}$, of ~3.3 K and a Berezinskii–Kosterlitz–Thouless (BKT) transition temperature $T_{BKT}$ of ~0.7 K (Discussion V [[39]]), under a direct laser excitation of 16 μW. Since non-resonant excitation may raise their temperature above that of the lattice, IXs trapped in the bright spots allow for more efficient IX cooling and offer a more viable pathway for achieving an exciton condensate. Our experiments suggest IXs confined in such traps have quantum degeneracy and superfluid temperatures of ~1.0 K and ~0.2 K, respectively, already accessible in an optical dilution refrigerator.

Nevertheless, to enhance the critical temperatures, it is desirable to further increase the IX densities. The upper limit of $n_{IX}$ imposed by the Mott transition is well above our experimental values, given their large binding energy (~81 meV). We estimate maximum quantum degeneracy $T_q^{max}$ and superfluid temperatures $T_{BKT}^{max}$ to be ~94 K and ~19 K, near the Mott transition (see Discussion V and **Fig. S14** [[39]]). Experimentally realizing such high densities necessitates further device optimization, including reducing gate current through resonant excitation and increasing the hBN gate thickness (**Fig. S7f**, sample D2). Another intriguing approach is to electrically inject IX, which may reduce heating in the bilayer system[10, 67, 68].

Crucially, disorder may disrupt the macroscopic coherence of the superfluid when its strength $\Delta$ exceeds the thermal energy at the critical temperature, $\Delta > k_B T_{BKT}$ [12]. Using the measured PL linewidth (~3.7 meV) as a proxy for the maximum disorder strength Δ, we find $\Delta$ already significantly lower than the thermal energies at $T_q^{max}$ and comparable to that at $T_{BKT}^{max}$. Another important consideration is the flavors of excitons, such as valley degeneracy and dark excitons. Although valley degeneracy in IXs would halve the quantum degeneracy temperature, leveraging the optical polarization of IXs shown here, possibly enhanced through resonant excitation, can substantially raise the condensate temperature. Likewise, dark IXs, having higher energies than bright ones in the intrinsic regime, might have a limited impact on the exciton condensate. In addition to quantum manipulation of bosonic particles, the electrical control and transport of robust interlayer excitons could form a basis for novel optoelectronics and valleytronics, such as excitonic transistors[19, 69] and manipulation of spin currents.

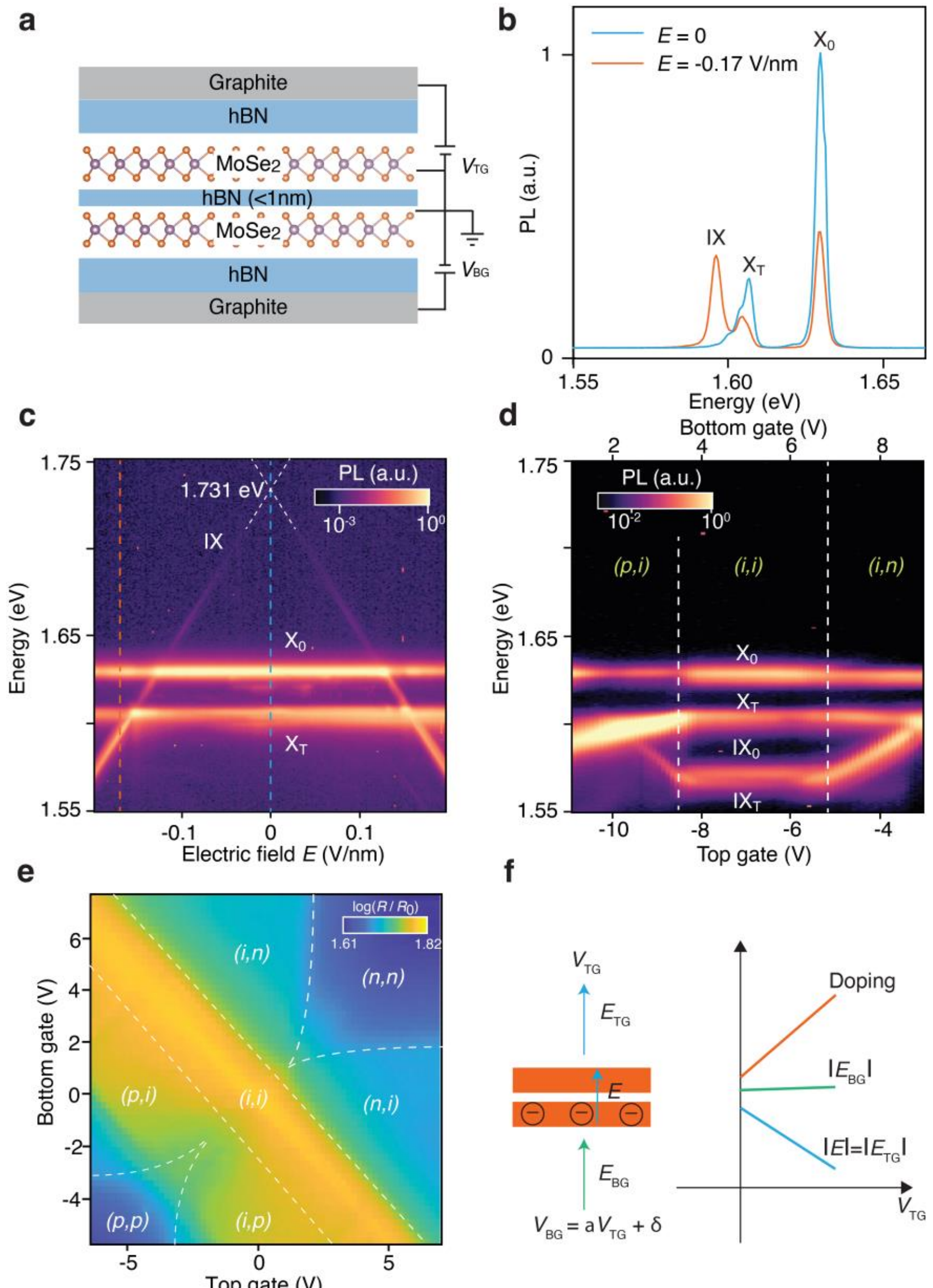

**Fig. 1 Electrical control of IX.** **a**, Schematic of the designed device. **b**, Representative PL spectrum under no electric field and -0.17 V/nm (negative defined as pointing upward). **c**, Electric field dependence of PL spectra when both layers are intrinsic. Dashed lines indicate where the spectra in **b** are collected. **d**, Doping dependence of PL spectra ($V_{BG} = 1.12V_{TG} + 12$ V). **e**, A 2D map of reflection contrast at $X_0$ as a function of top and bottom gates. **f**, Left: schematic of the electric field across the device under an asymmetric doping scan ($V_{BG} = 1.12V_{TG} + \delta$ and $\delta > 0$). A larger bottom gate voltage induces layer-polarized doping. Right: the magnitude of doping and *E*-field across the two layers versus $V_{TG}$.

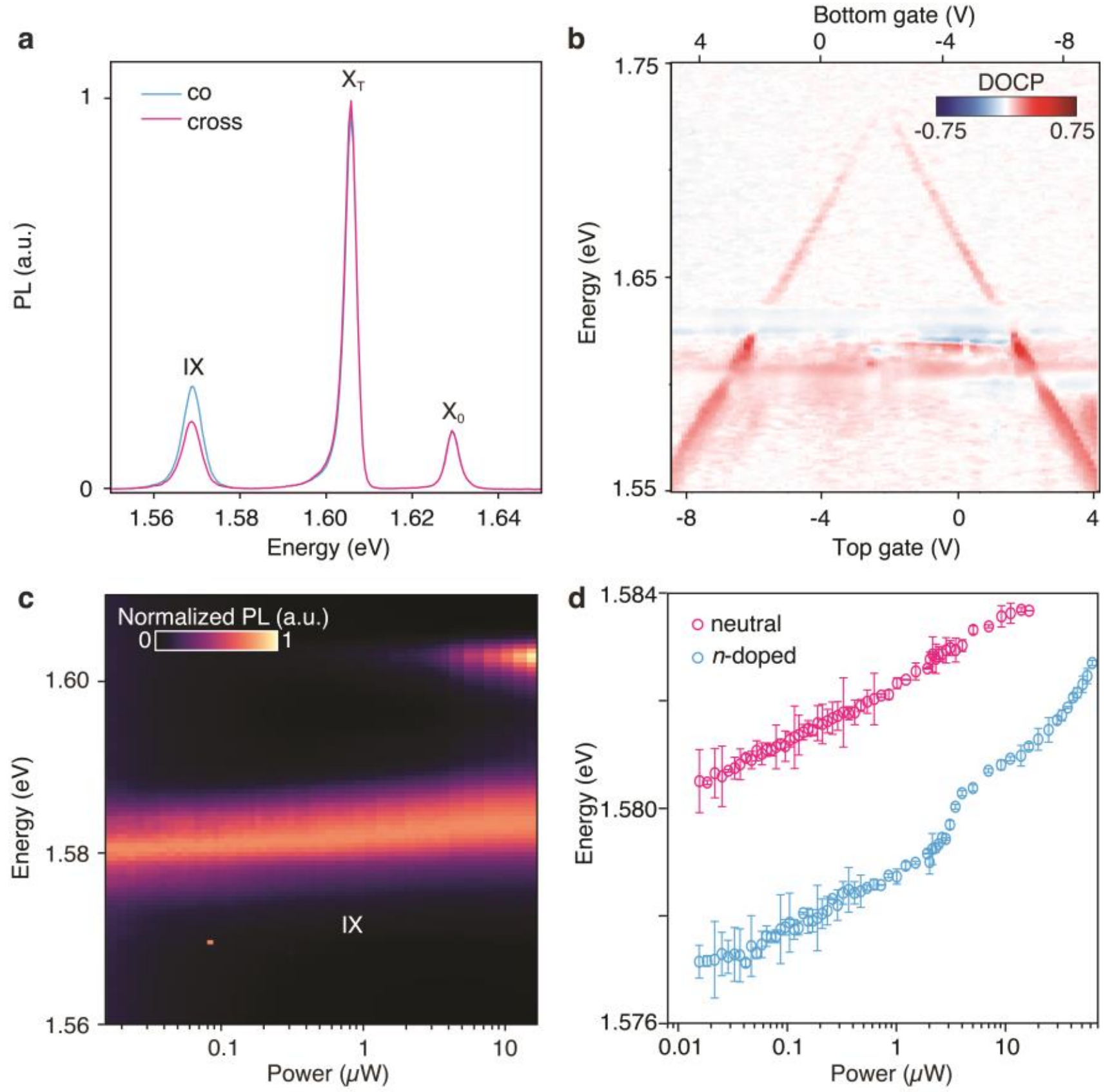

**Fig. 2 Valley polarization and dipolar interactions of IX. a**, Polarization resolved PL emission from $MoSe_2$/hBN/$MoSe_2$ under a circularly polarized laser excitation. **b**, DOCP as a function of the electric field under hole doping. **c**, Power-dependent PL emission from IX under an electric field of -0.2 V/nm in the intrinsic regime. A long-pass filter is used to block the intralayer exciton signals. **d**, Blueshift of the neutral and *n*-doped IX as a function of the laser power.

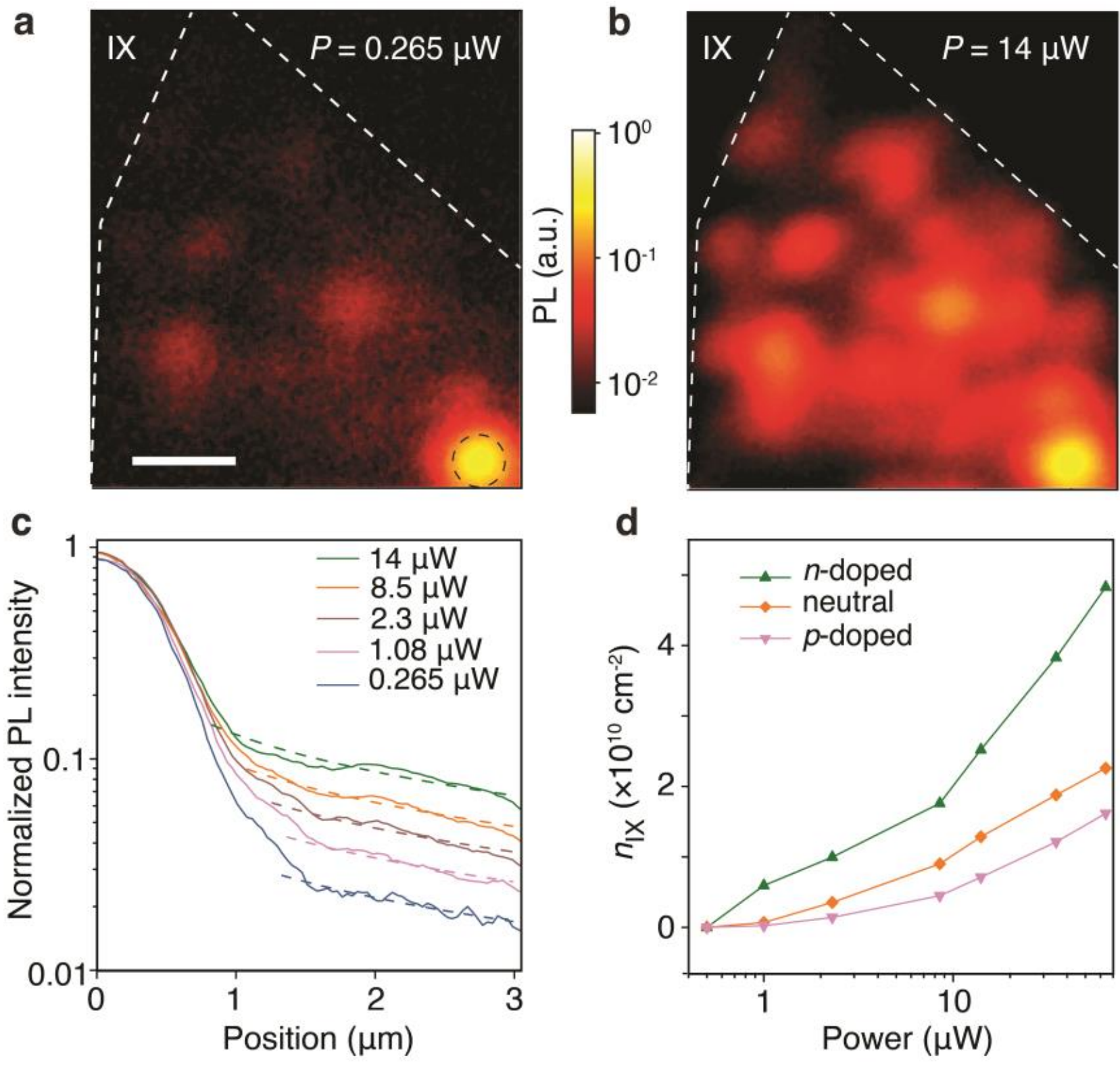

**Fig. 3 Spatial diffusion and trapping of IX. a-b**, Spatial map of IX emission under different excitation powers under an electric field of -0.2 V/nm, with fixed laser location (black dashed circle). Scale bar: 2 μm. **c**, Radially averaged IX PL intensity a function of the distance from laser center *r* under varying excitation powers. Dashed lines: fitted curves. **d**, Estimated IXs density in a natural trap as a function of laser power.

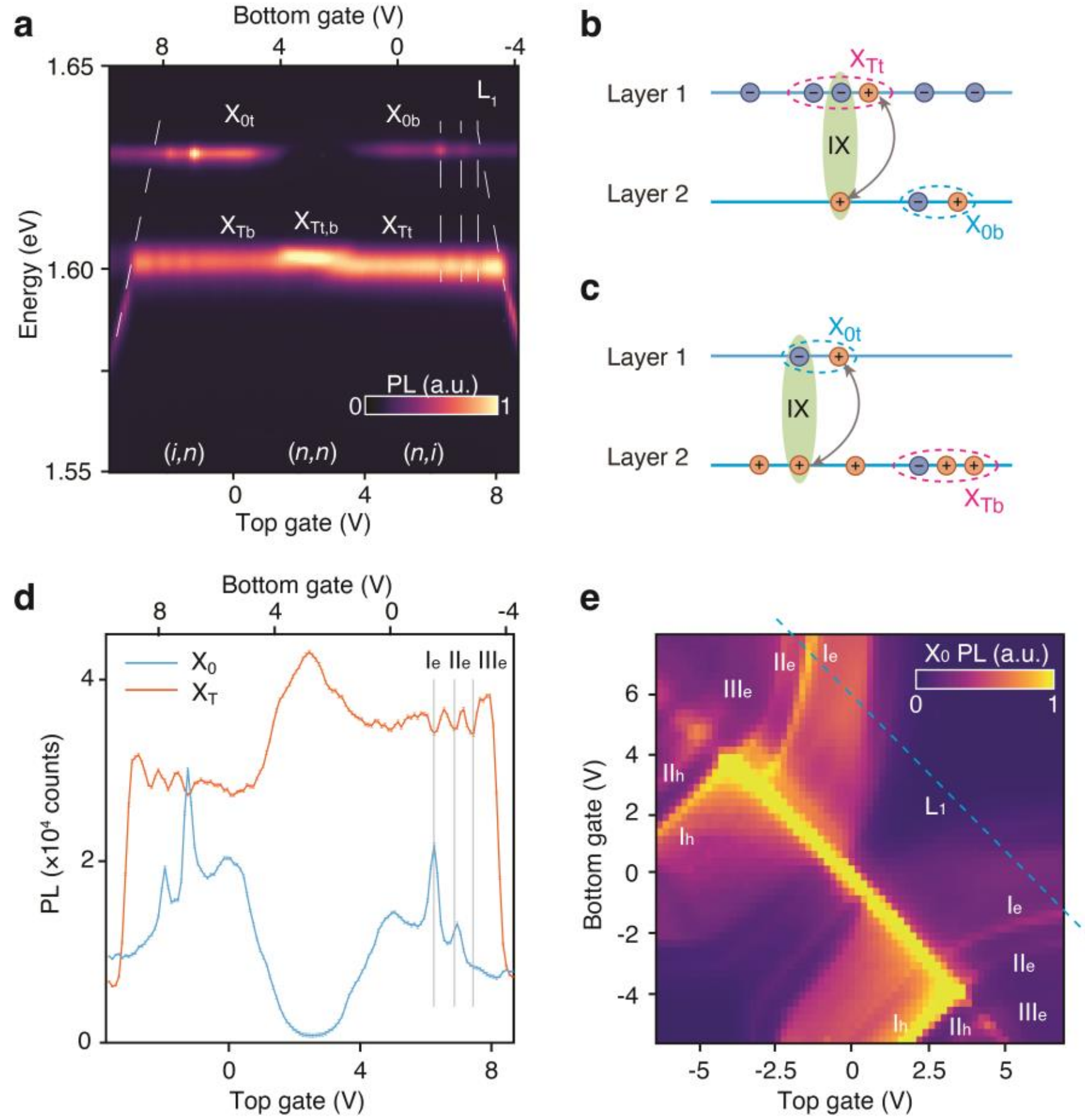

**Fig. 4 Coupling between intra- and inter-layer excitons.** **a**, PL spectra of a $MoSe_2$/hBN/$MoSe_2$ device as a function of electric field with a fixed electron doping. **b, c**, Hole tunneling facilitates (**b**) IX-$X_T$ coupling under electron doping and (**c**) IX-$X_0$ under hole doping. **d**, $X_0$ and $X_T$ intensities as a function of electric field. $I_e$, $II_e$, and $III_e$ donate three anomalies, corresponding to the vertical dashed lines in **a**. **e**, A 2d map of $X_0$. The dashed cyan line depicts the voltage conditions where **a** is collected.

**Acknowledgements:**

This research is primarily supported by the U.S. Department of Energy, Office of Science, Office of Basic Energy Sciences Early Career Research Program under Award No. DE-SC-0022885. The fabrication of samples is supported by the National Science Foundation CAREER Award under Award No. DMR-2145712. This research used Quantum Material Press (QPress) of the Center for Functional Nanomaterials (CFN), which is a U.S. Department of Energy Office of Science User Facility, at Brookhaven National Laboratory under Contract No. DE-SC0012704. K.W. and T.T. acknowledge support from the JSPS KAKENHI (Grant Numbers 20H00354, 21H05233 and 23H02052) and World Premier International Research Center Initiative (WPI), MEXT, Japan for hBN synthesis.

**Author contributions**

Y.Z. and L.Z. conceived the project. L.Z. fabricated the samples and performed the experiments. R.N., L.G., S.P., and H.J. assisted with sample fabrication. L.G. and R.N. helped with optical measurements. L.Z., M.X., R.N. and Y.Z. contributed to the data analysis and theoretical understanding. T.T. and K.W. provided hexagonal boron nitride samples. L.Z. and Y.Z. wrote the manuscript with extensive input from the other authors.

**References**

1. Butov L*, et al.* Stimulated scattering of indirect excitons in coupled quantum wells: signature of a degenerate Bose-gas of excitons. *Phys Rev Lett* **86**, 5608 (2001).
2. High AA*, et al.* Spontaneous coherence in a cold exciton gas. *Nature* **483**, 584-588 (2012).
3. Combescot M, Combescot R, Dubin F. Bose–Einstein condensation and indirect excitons: a review. *Rep Prog Phys* **80**, 066501 (2017).
4. Kohn W, Sherrington D. Two kinds of bosons and Bose condensates. *Reviews of Modern Physics* **42**, 1 (1970).
5. Rivera P*, et al.* Observation of long-lived interlayer excitons in monolayer MoSe2–WSe2 heterostructures. *Nature communications* **6**, 6242 (2015).
6. Fogler M, Butov L, Novoselov K. High-temperature superfluidity with indirect excitons in van der Waals heterostructures. *Nature communications* **5**, 4555 (2014).
7. Sivalertporn K, Mouchliadis L, Ivanov A, Philp R, Muljarov EA. Direct and indirect excitons in semiconductor coupled quantum wells in an applied electric field. *Physical Review B* **85**, 045207 (2012).
8. Chiaruttini F*, et al.* Trapping dipolar exciton fluids in GaN/(AlGa) N nanostructures. *Nano Lett* **19**, 4911-4918 (2019).
9. Wang Z*, et al.* Evidence of high-temperature exciton condensation in two-dimensional atomic double layers. *Nature* **574**, 76-80 (2019).
10. Xie M, MacDonald AH. Electrical reservoirs for bilayer excitons. *Phys Rev Lett* **121**, 067702 (2018).
11. Zeng Y, MacDonald A. Electrically controlled two-dimensional electron-hole fluids. *Physical Review B* **102**, 085154 (2020).
12. Wu F-C, Xue F, MacDonald A. Theory of two-dimensional spatially indirect equilibrium exciton condensates. *Physical Review B* **92**, 165121 (2015).

13. Berman OL, Kezerashvili RY. Superfluidity of dipolar excitons in a transition metal dichalcogenide double layer. *Physical Review B* **96**, 094502 (2017).
14. Unuchek D*, et al.* Valley-polarized exciton currents in a van der Waals heterostructure. *Nature nanotechnology* **14**, 1104-1109 (2019).
15. Rivera P*, et al.* Interlayer valley excitons in heterobilayers of transition metal dichalcogenides. *Nature nanotechnology* **13**, 1004-1015 (2018).
16. Rivera P*, et al.* Valley-polarized exciton dynamics in a 2D semiconductor heterostructure. *Science* **351**, 688-691 (2016).
17. Sun Z*, et al.* Excitonic transport driven by repulsive dipolar interaction in a van der Waals heterostructure. *Nature photonics* **16**, 79-85 (2022).
18. Yu L*, et al.* Observation of quadrupolar and dipolar excitons in a semiconductor heterotrilayer. *Nature materials*, 1-7 (2023).
19. Unuchek D*, et al.* Room-temperature electrical control of exciton flux in a van der Waals heterostructure. *Nature* **560**, 340-344 (2018).
20. Jauregui LA*, et al.* Electrical control of interlayer exciton dynamics in atomically thin heterostructures. *Science* **366**, 870-875 (2019).
21. Tagarelli F*, et al.* Electrical control of hybrid exciton transport in a van der Waals heterostructure. *Nature Photonics*, 1-7 (2023).
22. Paik EY*, et al.* Interlayer exciton laser of extended spatial coherence in atomically thin heterostructures. *Nature* **576**, 80-84 (2019).
23. Choi J*, et al.* Moiré potential impedes interlayer exciton diffusion in van der Waals heterostructures. *Science advances* **6**, eaba8866 (2020).
24. Yuan L*, et al.* Twist-angle-dependent interlayer exciton diffusion in WS2–WSe2 heterobilayers. *Nature materials* **19**, 617-623 (2020).
25. Wang J*, et al.* Diffusivity reveals three distinct phases of interlayer excitons in MoSe 2/WSe 2 heterobilayers. *Phys Rev Lett* **126**, 106804 (2021).
26. Weston A*, et al.* Atomic reconstruction in twisted bilayers of transition metal dichalcogenides. *Nature nanotechnology* **15**, 592-597 (2020).
27. Zhao S*, et al.* Excitons in mesoscopically reconstructed moiré heterostructures. *Nature nanotechnology*, 1-8 (2023).
28. Sung J*, et al.* Broken mirror symmetry in excitonic response of reconstructed domains in twisted MoSe2/MoSe2 bilayers. *Nature nanotechnology* **15**, 750-754 (2020).
29. Scuri G*, et al.* Electrically tunable valley dynamics in twisted WSe 2/WSe 2 bilayers. *Phys Rev Lett* **124**, 217403 (2020).
30. Zhang Y*, et al.* Every-other-layer dipolar excitons in a spin-valley locked superlattice. *Nature nanotechnology* **18**, 501-506 (2023).
31. Fowler-Gerace L, Zhou Z, Szwed E, Choksy D, Butov L. Transport and localization of indirect excitons in a van der Waals heterostructure. *arXiv preprint arXiv:230700702*, (2023).
32. Wietek E*, et al.* Non-linear and negative effective diffusivity of optical excitations in moir\'e-free heterobilayers. *arXiv preprint arXiv:230612339*, (2023).
33. Mahdikhanysarvejahany F*, et al.* Localized interlayer excitons in MoSe2–WSe2 heterostructures without a moiré potential. *Nature communications* **13**, 5354 (2022).
34. Andersen TI*, et al.* Excitons in a reconstructed moiré potential in twisted WSe2/WSe2 homobilayers. *Nature materials* **20**, 480-487 (2021).
35. Kim DS*, et al.* Electrostatic moiré potential from twisted hexagonal boron nitride layers. *Nature materials*, (2023).
36. Wang Z, Chiu Y-H, Honz K, Mak KF, Shan J. Electrical tuning of interlayer exciton gases in WSe2 bilayers. *Nano Lett* **18**, 137-143 (2018).
37. Shimazaki Y*, et al.* Strongly correlated electrons and hybrid excitons in a moiré heterostructure. *Nature* **580**, 472-477 (2020).

38. Schwartz I, *et al.* Electrically tunable feshbach resonances in twisted bilayer semiconductors. *Science* **374**, 336-340 (2021).
39. See Supplemental Material for additional details regarding experimental methods, discussions, other figures and references.
40. Chernikov A, *et al.* Exciton binding energy and nonhydrogenic Rydberg series in monolayer WS 2. *Phys Rev Lett* **113**, 076802 (2014).
41. Stier AV, *et al.* Magnetooptics of exciton Rydberg states in a monolayer semiconductor. *Phys Rev Lett* **120**, 057405 (2018).
42. Zhou Y, *et al.* Controlling excitons in an atomically thin membrane with a mirror. *Phys Rev Lett* **124**, 027401 (2020).
43. Van der Donck M, Peeters F. Interlayer excitons in transition metal dichalcogenide heterostructures. *Physical Review B* **98**, 115104 (2018).
44. Gu L, *et al.* Giant optical nonlinearity of Fermi polarons in atomically thin semiconductors. *arXiv preprint arXiv:230611199*, (2023).
45. Larentis S, *et al.* Large effective mass and interaction-enhanced Zeeman splitting of K-valley electrons in MoSe 2. *Physical Review B* **97**, 201407 (2018).
46. Zhang Y, *et al.* Direct observation of the transition from indirect to direct bandgap in atomically thin epitaxial MoSe2. *Nature nanotechnology* **9**, 111-115 (2014).
47. Glazov MM, *et al.* Exciton fine structure and spin decoherence in monolayers of transition metal dichalcogenides. *Physical Review B* **89**, 201302 (2014).
48. Yu T, Wu M. Valley depolarization due to intervalley and intravalley electron-hole exchange interactions in monolayer MoS 2. *Physical Review B* **89**, 205303 (2014).
49. Mai C, *et al.* Many-body effects in valleytronics: direct measurement of valley lifetimes in single-layer MoS2. *Nano Lett* **14**, 202-206 (2014).
50. Zhu C, *et al.* Exciton valley dynamics probed by Kerr rotation in WSe 2 monolayers. *Physical Review B* **90**, 161302 (2014).
51. Wang G, *et al.* Colloquium: Excitons in atomically thin transition metal dichalcogenides. *Reviews of Modern Physics* **90**, 021001 (2018).
52. Wu F, Lovorn T, MacDonald A. Theory of optical absorption by interlayer excitons in transition metal dichalcogenide heterobilayers. *Physical Review B* **97**, 035306 (2018).
53. Schaibley JR, *et al.* Valleytronics in 2D materials. *Nature Reviews Materials* **1**, 1-15 (2016).
54. Sushko A, *et al.* Asymmetric photoelectric effect: Auger-assisted hot hole photocurrents in transition metal dichalcogenides. *Nanophotonics* **10**, 105-113 (2020).
55. Yoshioka D, MacDonald AH. Double quantum well electron-hole systems in strong magnetic fields. *J Phys Soc Jpn* **59**, 4211-4214 (1990).
56. Sun D, *et al.* Observation of rapid exciton–exciton annihilation in monolayer molybdenum disulfide. *Nano Lett* **14**, 5625-5629 (2014).
57. Andreev S. Spin-orbit-coupled depairing of a dipolar biexciton superfluid. *Physical Review B* **103**, 184503 (2021).
58. Kulig M, *et al.* Exciton diffusion and halo effects in monolayer semiconductors. *Phys Rev Lett* **120**, 207401 (2018).
59. Glazov M. Phonon wind and drag of excitons in monolayer semiconductors. *Physical Review B* **100**, 045426 (2019).
60. Pisoni R, *et al.* Absence of interlayer tunnel coupling of K-valley electrons in bilayer MoS 2. *Phys Rev Lett* **123**, 117702 (2019).
61. Liu G-B, Xiao D, Yao Y, Xu X, Yao W. Electronic structures and theoretical modelling of two-dimensional group-VIB transition metal dichalcogenides. *Chem Soc Rev* **44**, 2643-2663 (2015).
62. Yu H, Liu G-B, Tang J, Xu X, Yao W. Moiré excitons: From programmable quantum emitter arrays to spin-orbit–coupled artificial lattices. *Science advances* **3**, e1701696 (2017).
63. Jin C, *et al.* Interlayer electron–phonon coupling in WSe2/hBN heterostructures. *Nature Physics* **13**, 127-131 (2017).

64. Serrano J*, et al.* Vibrational properties of hexagonal boron nitride: inelastic X-ray scattering and ab initio calculations. *Phys Rev Lett* **98**, 095503 (2007).
65. Soubelet P, Bruchhausen AE, Fainstein A, Nogajewski K, Faugeras C. Resonance effects in the Raman scattering of monolayer and few-layer MoSe 2. *Physical Review B* **93**, 155407 (2016).
66. Chow CM*, et al.* Phonon-assisted oscillatory exciton dynamics in monolayer MoSe2. *npj 2D Materials and Applications* **1**, 33 (2017).
67. Qi R*, et al.* Perfect Coulomb drag and exciton transport in an excitonic insulator. *arXiv preprint arXiv:230915357*, (2023).
68. Nguyen PX*, et al.* Perfect Coulomb drag in a dipolar excitonic insulator. *arXiv preprint arXiv:230914940*, (2023).
69. High AA, Novitskaya EE, Butov LV, Hanson M, Gossard AC. Control of exciton fluxes in an excitonic integrated circuit. *Science* **321**, 229-231 (2008).